\documentclass[a4paper,10pt,aps,prd,showpacs,preprintnumbers,twocolumn,nofootinbib,superscriptaddress]{revtex4-2}

\usepackage[colorlinks=true,
            linkcolor=blue,
            citecolor=blue,
            urlcolor=blue]{hyperref}

\def \be {\begin{equation}}
\def \ee {\end{equation}}
\def \dd {\mathrm{d}}

\def \l {\left}
\def \r {\right}

\def \bs {\boldsymbol}

\newcommand{\e}[1]{_{\rm #1}}

\newcommand{\beq}{\begin{equation}}
\newcommand{\eeq}{\end{equation}}
\newcommand{\bea}{\begin{eqnarray}}
\newcommand{\eea}{\end{eqnarray}}


\newcommand\ees{\end{eqnarray}}
\newcommand\bees{\begin{eqnarray}}

\def\dd{\mathrm{d}}

\usepackage{amsfonts}
\usepackage{lettrine}
\usepackage{graphicx}
\usepackage{booktabs}
\usepackage{comment}
\usepackage[usenames,dvipsnames]{xcolor}
\usepackage{amsmath}
\usepackage{url}
\usepackage{gensymb}
\usepackage{soul}
\usepackage[utf8]{inputenc}
\usepackage{orcidlink}
\usepackage[normalem]{ulem}
\usepackage{empheq}

\def\dd{\mathrm{d}}

\begin{document}

\title{On the origin of Milky Way binary black holes with LISA}

\author{Charles Dalang\orcidlink{0000-0002-7373-6903}}
\email{charles.dalang@ens.fr}

\affiliation{Institut Philippe Meyer, Département de Physique,\\ École Normale Supérieure (ENS), Université PSL, Paris, France}

\author{Enrique Gazta$\bs{\tilde{\hbox{n}}}$aga\orcidlink{
0000-0001-9632-0815}}
\affiliation{Institute of Cosmology and Gravitation, University of Portsmouth, \\
Burnaby Road, Portsmouth PO1 3FX, United Kingdom}
\affiliation{Institute of Space Sciences (ICE, CSIC), 08193 Barcelona, Spain}
\affiliation{Institut d'
Estudis Espacials de Catalunya (IEEC), 08034 Barcelona, Spain}

\date{\today }


\begin{abstract} 
Primordial black holes (PBHs) may constitute a fraction of dark matter, but distinguishing them from black holes of stellar origin remains a major challenge. We propose a new method based on the three-dimensional localization of Galactic Binary Black Holes (BBHs) by the Laser Interferometer Space Antenna (LISA). While stellar-origin BBHs are expected to be associated with the stellar distribution of the Milky Way, PBHs may trace its more extended dark matter halo. We introduce a spatial overlap ratio that compares the gravitational-wave localization of each binary with the normalized stellar and dark matter density profiles. Using a Fisher-matrix forecast and 1000 Monte Carlo realizations of a 10-year LISA mission, we find that detecting three BBHs with an overlap ratio $\ln B<0$ occurs in approximately 99\% of simulations where BBHs trace dark matter, but in fewer than 0.1\% of simulations where they follow the stellar distribution, even when including a stellar halo containing 3\% of the Galactic stellar mass. These results suggest that the localization of only a few Galactic BBHs could provide strong evidence for a primordial population, offering an independent test of their possible contribution to dark matter.
\end{abstract}

\maketitle

\section{Introduction}

\lettrine{P}{}rimordial Black Holes (PBHs) may form from the collapse of large density inhomogeneities in the early Universe, particularly
during the radiation-dominated era \cite{Zeldovich:1967lct,Hawking:1971ei,Carr:1974nx}. Other formation mechanisms have also been proposed, including phase transitions and topological defects
\cite{CarrIovino2026}. They may also arise as relics of a preceding cosmological bouncing phase \cite{2026PhRvD.113d3544G}. They could seed supermassive black holes at the centres of galaxies \cite{Carr:2018rid} and may constitute a fraction of the dark matter (DM) density today \cite{CarrKohri2021,CarrIovino2026}.Although their abundance is constrained by gravitational lensing \cite{Macho:2000nvd, EROS-2:2006ryy, Niikura:2017zjd}, dynamical effects \cite{Brandt:2016aco}, accretion \cite{Ali-Haimoud:2016mbv, Serpico:2020ehh} and gravitational-wave observations \cite{Ali-Haimoud:2017rtz,
Hutsi:2020sol} (see Refs.~\cite{Carr:2020gox, Green:2020jor} for reviews),
a subdominant population remains of considerable cosmological interest. Therefore, any unambiguous detection of a PBH would have wide implications for astrophysics and cosmology.

Black holes of stellar origin, formed through the evolution and collapse of massive stars, provide a competing explanation for the observed population of binary black holes (BBHs). Distinguishing between these formation channels is therefore a central question in gravitational-wave astronomy \cite{Sasaki2016,Sasaki2018}. Discriminating parameters include the component mass and spin distributions, the redshift dependence of the merger rate, orbital eccentricity and possible tidal effects \cite{Franciolini:2021xbq}. However, these observables may exhibit degeneracies between primordial and astrophysical formation scenarios, motivating the search for independent and complementary diagnostics.

The spatial distribution of BBHs provides such an opportunity. Stellar origin BBHs are expected to be associated with the stellar populations of their host galaxies, whereas PBHs constituting a fraction of DM may approximately trace the dark matter halo. These populations have markedly different spatial distributions in the Milky Way: most stars are concentrated in the Galactic disc and bulge, while the dark matter halo extends to much larger Galactocentric distances. A BBH sufficiently well localized in the outer Galaxy could therefore favour a primordial origin over
a conventional stellar origin. The main astrophysical complication is the existence of a stellar halo, together with the possible migration of stellar-origin BBHs away from their formation sites. These effects must be included when assessing the significance of such a detection.

The Laser Interferometer Space Antenna (LISA) offers a particularly interesting opportunity to explore this possibility. Stellar-mass BBHs in the Milky Way can emit gravitational waves in the LISA frequency band long before their final merger \cite{Christian:2017pmf,Lamberts:2018}. The orbital motion of LISA and the associated modulation of the gravitational-wave signal can constrain both the sky position and luminosity distance of these sources. Their three-dimensional localization can therefore provide information about their Galactic environment, complementary to their intrinsic binary properties.

In this work, we propose an alternative diagnostic for identifying candidate PBHs associated with DM, based on the ratio of the overlaps between the three-dimensional GW localization distribution and the stellar and DM density profiles of the Milky Way. We focus on Milky Way BBHs in the LISA band and investigate the ability of this overlap ratio to distinguish stellar-origin BBHs from a population tracing the Galactic DM halo. We estimate the number of Galactic BBHs in the relevant frequency range based on current merger rates from the LIGO--Virgo--KAGRA (LVK) collaboration.
We generate catalogues of Galactic BBHs detectable
by LISA by selecting those which have a sufficiently
large signal-to-noise ratio (SNR). We perform a Fisher forecast to determine the uncertainties on their three-dimensional localization and compute their stellar and DM halo overlaps under the two formation hypotheses.
Finally, we run $N_{\rm sim}=1000$ simulations to quantify the probability of identifying one or several BBHs whose locations are difficult to reconcile with a stellar origin, and the corresponding false-alarm probability under the stellar hypothesis. We also investigate how these results depend on the assumed contribution of the stellar halo.

This work is structured as follows. In Sec.\,\ref{sec:MW_BBH}, we estimate the number of Milky Way BBH from current LVK merger rates. In Sec.\,\ref{sec:MW_stellar_vs_PBH_density}, we describe the stellar and DM density profiles from which we sample the 3D positions of the BBH population.
In Sec.\,\ref{sec:GW_events_LISA}, we describe the LISA detectability calculation and Fisher forecast used to estimate the localization uncertainties. In Sec.\,\ref{sec:Stellar_Overlap}, we introduce the stellar and DM halo overlaps and their ratio as a spatial discrimination statistic.
We present and discuss the results in Sec.\,\ref{sec:Results} and conclude in Sec.\,\ref{sec:Conclusion}.

\section{Milky Way Binary Black Holes} \label{sec:MW_BBH}

It is expected that there are stellar mass Binary Black Holes (BBH) in the Milky Way \cite{Christian:2017pmf}. In this section, we estimate the expected number of BBH that fall in the LISA band $[10^{-4}, 10^{-1}]$ Hz from the observed GW merger rate. To this end, we adapt some of the results of \cite{Christian:2017pmf} to unequal mass binaries. We consider BBHs that have similar properties as those observed by LVK, in the $[4,100] M_\odot$ range. The number of BBH of masses $m_1,m_2$ with semi-major axis $a$ smaller than $a\e{max} = 10 R_\odot$ is
\begin{align}
N(a<a\e{max})= R\e{MW} \cdot t(m_1,m_2)
\end{align}
where $t$ is the time that it takes for such a BBH to merge and $R\e{MW}$ is the Milky Way merger rate. This time can be computed from the relation
\begin{align}
\frac{\dd a}{\dd t} = - K a^{-3}\,, \label{eq:dadt}
\end{align}
where 
\begin{align}
K = \frac{64 G^3}{5c^5} m_1m_2 (m_1+ m_2)\,.
\end{align}
In words, the more massive the BBH, the faster their orbit shrinks. 
The relation \eqref{eq:dadt} follows essentially from the quadrupole formula. The time that a BBH spends below a semi-major axis $a\e{max}$ can be obtained by integrating Eq.\,\eqref{eq:dadt}
\begin{align}
t(m_1,m_2) = \frac{a\e{max}^4}{4 K}
\end{align}
One can also estimate the number of BBH below a certain semi-major axis by weighting by the number of BBH with given primary and secondary masses $m_1> m_2$
\begin{align}
N(a\leq a\e{max}) & = R\e{MW} \int_{m\e{min}}^{m\e{max}} \dd m_1 \Phi(m_1) \\
& \quad  \times\int_{m\e{min}}^{m_1} \dd m_2  \Phi(m_2|m_1) t(m_1,m_2)\,.
\end{align}
The Milky Way merger rate $R\e{MW}$ can be estimated from the LVK merger rate $R\e{LVK} = 100\,R_{100}$\,Gpc$^{-3}$\,yr$^{-1}$, with current constraints $R\e{LVK} \simeq 27$-$49$\,Gpc$^{-3}$\,yr$^{-1}$ at $90\%$ credible level
(at $z=0.2$) \cite{LIGOScientific:2026ctl}. For example, if all mergers occur in galaxies of Milky Way mass, whose number density is $\rho\e{MW} \simeq (0.5$--$1)\times10^{-2}\,h^3$\,Mpc$^{-3}$
\cite{Baldry_2020,Montero-Dorta:2008klw}, one obtains, for $h=0.7$,
\begin{align}
R\e{MW} = \frac{R\e{LVK}}{\rho\e{MW}} \simeq (3\text{--}6)\times10^{-5}\,R_{100}\,\hbox{yr}^{-1}\,. \label{eq:RMW_bracket}
\end{align}
This is an upper estimate, since lower-mass galaxies may also host mergers. However, it presents the advantage of being channel-formation independent. If one assumes that the BBH's detected from LVK come from stars, one may weight by stellar mass
\begin{align}
R\e{MW}^* = R\e{LVK}\,\frac{M_*}{\rho_*} \simeq 1.7\times10^{-5}\,R_{100}\,\hbox{yr}^{-1}\,,
\end{align}
where $M\e{MW,*} \simeq 5 \times10^{10}M_\odot$ is the Milky Way stellar mass \cite{Cautun_2020} and $\rho_* \simeq 3\times10^{8}M_\odot$\,Mpc$^{-3}$
is the local cosmic stellar mass density \cite{Driver:2022vyh}. If instead, one weights by DM density and mass, one obtains
\begin{align}
R\e{MW}^{\rm{PBH}} = R\e{LVK}\,\frac{M\e{DM}}{\rho\e{DM}} \simeq 3.1\times10^{-6}\,R_{100}\,\hbox{yr}^{-1}\,,
\end{align}
where $M\e{DM} = 0.97 \cdot 10^{12} M_\odot $ is the MW DM mass \cite{Cautun_2020} and $\rho\e{DM} \simeq 2.2\times10^{-27}$\,kg\,m$^{-3}$ is the cosmic average DM density \cite{Planck:2018vyg}. Therefore, the conversion of the cosmological BBH merger rate into a Galactic rate may depend on the assumed relation between merger activity and the properties of the host galaxy. Weighting by stellar mass gives a substantially larger Milky Way rate than weighting by DM mass, reflecting the fact that the Milky Way has a stellar-to-total mass ratio considerably above the cosmic average. The former scaling may be more appropriate for stellar-origin BBHs, while the latter could provide a useful benchmark for primordial populations tracing dark matter, provided their
merger rate per unit dark matter mass is approximately universal. These estimates are comparable with the population-synthesis prediction of $\sim10^{-5}$\,yr$^{-1}$ of \cite{Lamberts:2018}. While recognizing that the actual Galactic rate may depend strongly on the formation channel, it may also depend on metallicity and star-formation history \cite{Lamberts:2018}. Therefore, we consider for simplicity an optimistic scenario in which
\begin{align}
R\e{MW} &= 3\times10^{-5}\,\hbox{yr}^{-1}\,,\label{eq:RMW_optimistic}
\end{align}
falling at the upped end of the bracket allowed by \eqref{eq:RMW_bracket}. Since our results scale linearly with $R\e{MW}$, it is relatively easy to translate those to a more pessimistic scenario.

\begin{figure*}
    \centering
    \includegraphics[width=1.0\linewidth]{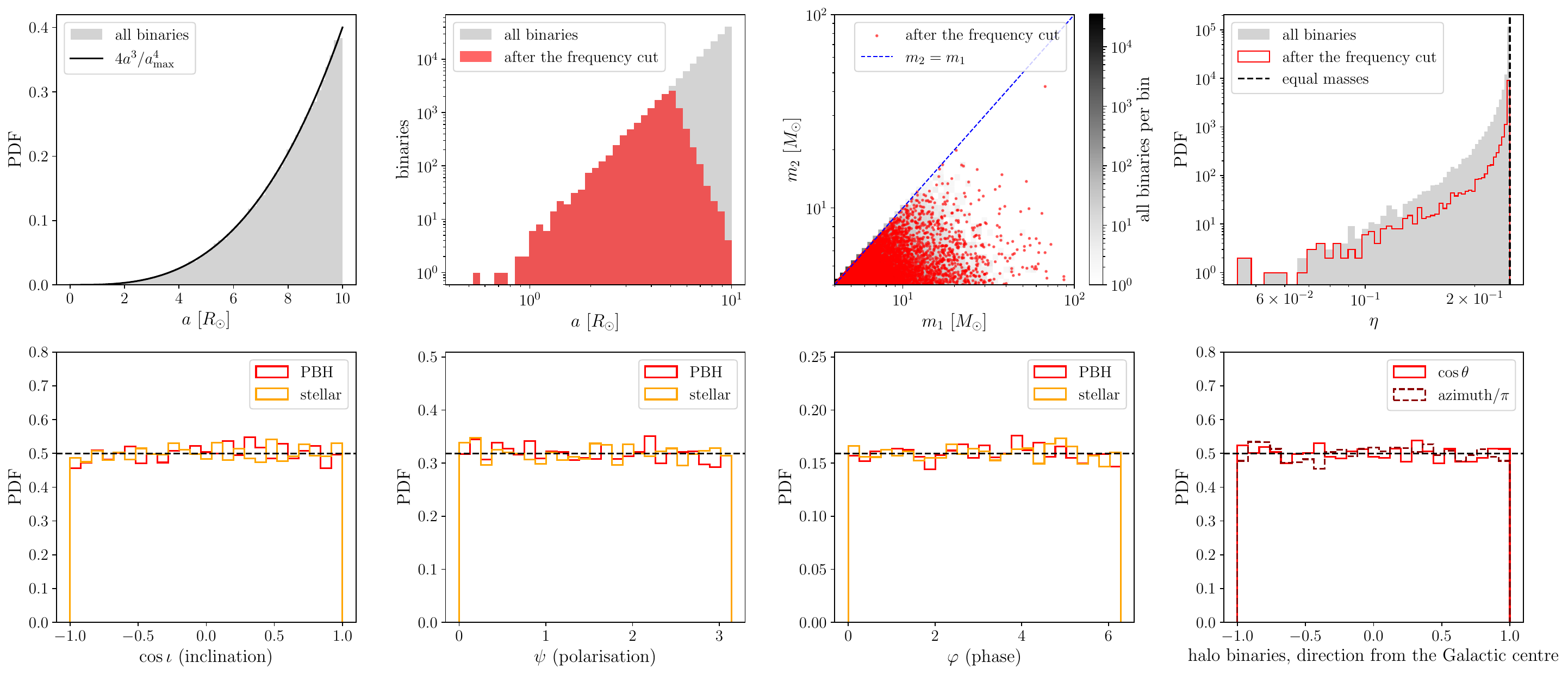}
    \caption{Properties of the simulated BBH for one realization. We present the semi-major axis $a$ distribution from the BBH before and after the frequency cut, the mass $m_1$ and $m_2$ of the component masses together with the mass ratio $\eta = m_1 m_2/M^2$ where $M = m_1 + m_2$, before and after the frequency cut. We also present the inclination $\cos \iota$, polarization $\psi$, coalescence phase $\varphi$ angle distribution together with the polar and azimuthal angular coordinates based on spherical coordinated centered on the MW center. }
    \label{fig:check_parameters}
\end{figure*}

The MW BBH have internal properties, which we sample using the following distributions. For the BBH masses, we adopt $\Phi(m_1) \propto m_1^{-n_1}$ in the interval $m_1\in[m\e{min},m\e{max}]$ and $\Phi(m_2|m_1) \propto m_2^{-n_2}$ in $m_2\in[m\e{min},m_1]$. For $m\e{min} = 4 M_\odot$, $m\e{max} = 100 M_\odot$ and $n_1 = n_2 = 2.5$, a simplified choice broadly consistent with the results from \cite{KAGRA:2021duu}, this gives approximately
\begin{align}
N(a\leq a\e{max}) \simeq 1.5 \cdot 10^5  \l(\frac{a\e{max}}{10 R_\odot} \r)^4\,.
\end{align}
The parameters left to describe the internal parameters of BBH are angles such as the inclination angle $\cos \iota$, polarization angle $\psi$ and coalescence phase $\phi$, which are presented together with the masses, semi-major axis and direction from the MW center distribution in Fig.\,\ref{fig:check_parameters}. In the following, we describe the mass density of the Milky Way to distribute these BBH in the galaxy, following either the stellar distribution or the dark matter distribution.

\section{Milky Way stellar and PBH density}\label{sec:MW_stellar_vs_PBH_density}

\begin{figure}
    \centering
    \includegraphics[width=1.0\linewidth]{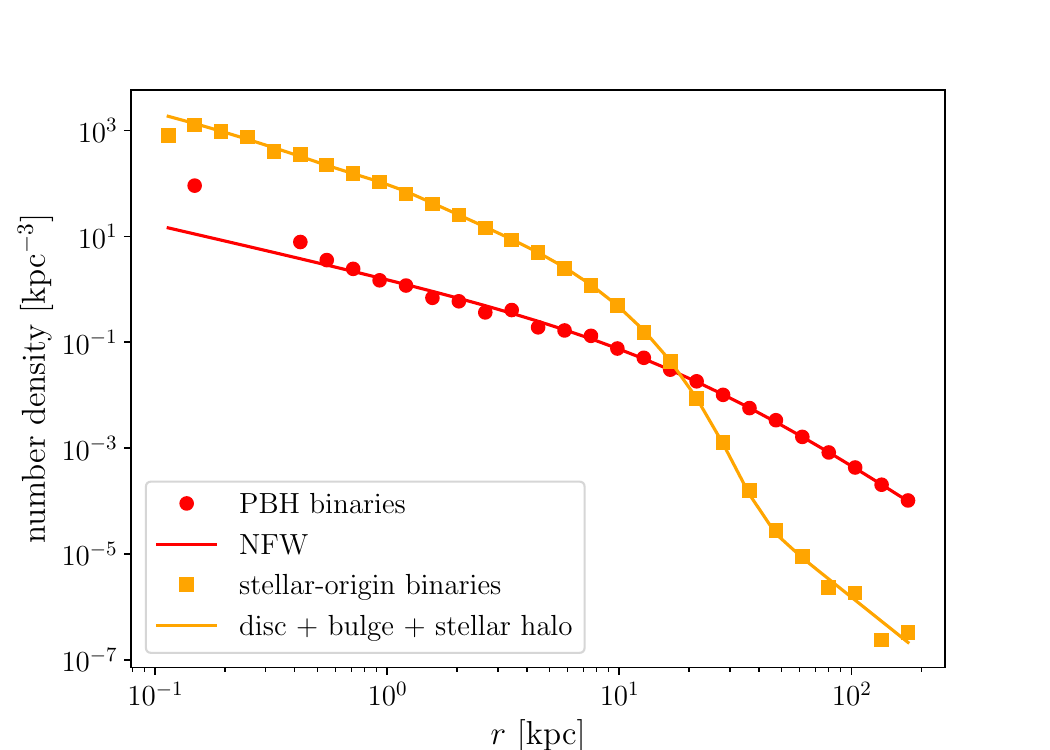}
    \caption{Radial density of BBH if they are of primordial or stellar origin. The stellar halo fraction here is $f\e{SH} = 0.03$. Without it, the number density of stellar-origin BBH would drop exponentially around 10kpc. }
    \label{fig:Radial_density}
\end{figure}

We aim to use the 3D localization of the BBH to tell wether they are of stellar or primordial origin. Indeed, if they are of stellar origin, they should somehow trace the stellar density field. In contrast, if the BBH are of primordial origin, they are not expected to trace the stellar density field, but may contribute to some fraction of the dark matter. In that case, they are expected to follow a Navarro-Frank-White (NFW) profile \cite{Navarro:1995iw}, which extends further out the stellar galaxy, as may be seen in Fig.\,\ref{fig:Radial_density}. If one can localize sufficiently well a BBH outside the regions with significant stellar density, this may be taken as evidence against the stellar origin of that BBH, although the presence of the stellar halo forms an obstacle to that evidence being a smoking gun.

The formation of BBH is very complicated and is the subject of BBH population synthesize codes \cite{Fragos:2022jik,Andrews:2024saw,Lamberts:2018}. Here instead, we focus on a simple model of the Milky Way density with a stellar bulge, disc and stellar halo, together with an NFW profile for dark matter. We use a simple galactic model with parameters for the densities taken from \cite{McMillan_2016} and normalize those using results from \cite{Cautun_2020}. The bulge is described by
\begin{align}
\rho_b(R,z) \propto \frac{1}{(1+r'/r_0)^\alpha}\hbox{exp}\l( -\l(\frac{r'}{r\e{cut}}\r)^2 \r)
\end{align}
with $r' = \sqrt{R^2 + (z/q)^2}$, with $\alpha= 1.8$, $r_0 = 0.075$ kpc, $r\e{cut} =2.1$ kpc, $q = 0.5$. We consider only the thick stellar disc described by
\begin{align}
\rho_d(R,z) \propto \frac{1}{2 z_d} \hbox{exp}\l( -\frac{|z|}{z_d} - \frac{R}{R_d}\r)
\end{align}
with $z\e{d;thick} = 900$pc, $R\e{d;thick} = (3.6 \pm 0.72)$ kpc and neglect the thin disc, which for our purpose is a conservative assumption. We also consider a stellar halo which amounts to $0.1-3\%$ of the total stellar mass of the Milky Way and which extends to $200$kpc \cite{Cooper_2010}. We parametrize the stellar halo in the standard way \cite{Amina:2008}
\begin{align}
\rho\e{SH}(x,y,z) \propto \l( x^2 + \frac{y^2}{p^2} + \frac{z^2}{q^2} \r)^{-n/2}
\end{align}
and take $n = 3.5$, $p = 1$, $q= 0.6$ and apply a strict upper cutoff of $200$kpc \cite{Morrison_2000}. The Dark Matter halo has the following shape
\begin{align}
\rho\e{NFW}(r) \propto \frac{1}{x^\gamma (1+x)^{3-\gamma}}\,.
\end{align}
with $x = r/r_h$ and $\gamma = 1$. We take the NFW halo turnover to be $r\e{h} = 20$ kpc and apply a cutoff such that $r< r\e{cut} = 200$ kpc. We assume that the Sun is located $8.2$ kpc away from the center of the MW \cite{2016ARA&A..54..529B} and on the $(x-y)$ plane. Note that at large galactic radius, the difference in falloffs with radius between the stellar halo and the NFW halo is $\rho\e{SH}/\rho\e{NFW}(r\gg R\e{d;thick}) \sim r^{-1/2}$. The matter densities are normalized such that integrated over a sphere of radius of 200 kpc, $M\e{bulge}/M\e{disc} = 0.1$. and $M_* = M\e{bulge}+M\e{disc} = 5.04 \cdot10^{10} M_\odot$ \cite{Cautun_2020}. We vary the stellar halo mass fraction in the interval $f\e{SH} \equiv M\e{SH}/M_* = [0.001,0.03]$. We assume that the DM halo mass density integrates to $M\e{DM} = 0.97 \cdot 10^{12} M_\odot$ \cite{Cautun_2020}. For a given set of GW internal parameters, we perform two simulations; one where the BBH are sampled from the stellar density, simulating stellar BBH, another one where the BBH are sampled from the NFW profile, simulating DM PBH. This simple MW model allows us to sample the BBH population in a way which is proportional to the stellar and DM density and to compare their spatial distributions.

Importantly, it neglects sophisticated astrophysics which may enhance the contribution from metal-poor stars in the stellar halo, to the stellar BBH population \cite{Lamberts:2018}. It also neglects natal kicks which are expected to displace the positions of hierarchical BBH mergers. Furthermore, this simplistic geometry allows to simulate the positions of BBH, neglecting the sophisticated geometry of the MW, for which data exists from GAIA \cite{Gaia:2023fqm}, for example. Nonetheless, we take this as a proof of concept and a more detailed prescription of the MW stellar BBH population including the effects of metallicity, natal kicks and satellite galaxies may be included in the method as an input without any major obstacle.

\section{GW events observable by LISA} \label{sec:GW_events_LISA}
Having fixed the 3D position of these BBH with their internal properties, one may estimate how many of these GW sources can LISA detect and with what uncertainties on their 3D positions. We first apply naive frequency and SNR cuts. We then use \texttt{fastgb} \cite{fastgb} to evaluate their SNR's and the uncertainties on their distance $d_L$ and solid angle $\Delta \Omega_{90}$.

\begin{figure}
    \includegraphics[width=1.0\linewidth]{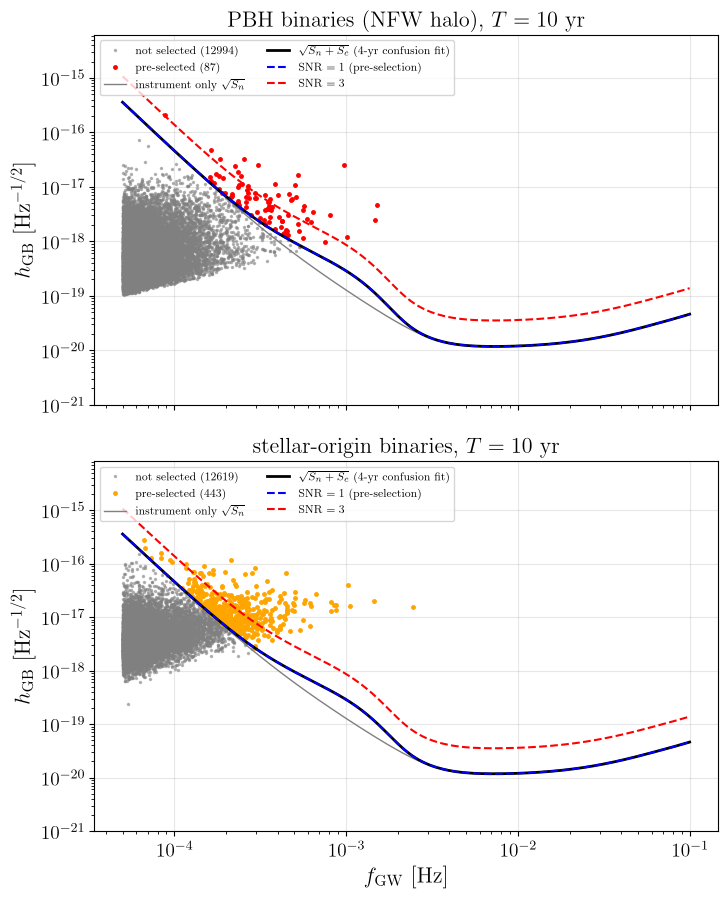}
    \caption{We present the SNR cuts that we apply to the two populations of BBH. The small bump in the sensitivity curve around $10^{-3}$Hz comes from confusion noise from galactic binary white dwarfs.}
    \label{fig:preselection}
\end{figure}

We first apply a rough frequency cut $0.5\cdot 10^{-5}\,\hbox{Hz}<f\e{gw}<0.1$ Hz and compute a rough SNR, by estimating a spectral density for nearly stationary galactic binaries of \cite{Robson:2018ifk}
\begin{align}
h\e{GB}(f) = \sqrt{\frac{4 T}{5}} h_c(f)\,,\label{eq:hGB}
\end{align}
which reads in [Hz$^{-1/2}$] and the characteristic strain amplitude is defined as
\begin{align}
h_c(f) = \frac{4 (G \mathcal{M}_c)^{5/3} (\pi f)^{2/3}}{c^4 D_L}\,. \label{eq:hc}
\end{align}
Indeed, the SNR scales with the mission duration time $T$ (which we take to be 10 years). The quantity defined in Eq.\,\eqref{eq:hGB} can directly be compared with an approximate sensitivity curve of LISA $S_n(f)$, for which we use Eq.\,(13) of \cite{Robson:2018ifk} together with confusion noise from galactic binary white dwarfs from their Eq.\,(14). The SNR for stationary sources is well approximated by the ratio $\rho = h\e{GB}/\sqrt{S_n(f)}$. We present the signals and sensitivity curve of LISA for the two populations in Fig.\,\ref{fig:preselection}. We use this approximate, sky- and orientation-averaged SNR to select sources with $\rho>1$ for a more detailed analysis with the galactic binary waveform generator \texttt{fastgb} \cite{fastgb}, as described below.

Each binary is modeled as a quasi-circular, slowly inspiralling source with gravitational-wave frequency
$f(t) = f_0 + \dot f\, t$, where, at leading quadrupole order,
$\dot f = \frac{96}{5}\pi^{8/3} (G\mathcal{M}/c^3)^{5/3} f_0^{11/3}$ \cite{Maggiore:2007ulw}, and with strain amplitude $h_c$ as given in Eq.\,\eqref{eq:hc} and phase $\Phi$ so that $h_+ = h_c(1+\cos^2\iota)\cos\Phi/2$ and $h_\times = h_c \cos\iota\sin\Phi$. The LISA response is computed with \texttt{fastgb}
\citep{Cornish:2007if}, which returns the frequency-domain Time Delay Interferometry (TDI) observables
$X,Y,Z$ including the amplitude modulation from the rotation of the constellation and the Doppler modulation from its orbit around the Sun \cite{Martens:2021phh}. We assume equal-arm orbits, which allows to form the noise-orthogonal channels $A = (Z - X)/\sqrt{2}$, $E = (X - 2Y + Z)/\sqrt{6}$ and
$T = (X + Y + Z)/\sqrt{3}$ \citep{Prince:2002hp}. The linear-chirp approximation is valid for almost all the sources considered here, except for about 2 per thousand detected sources, for which a phase shift from higher order correction is above 1\,rad.

\begin{figure}
    \includegraphics[width=1.0\linewidth]{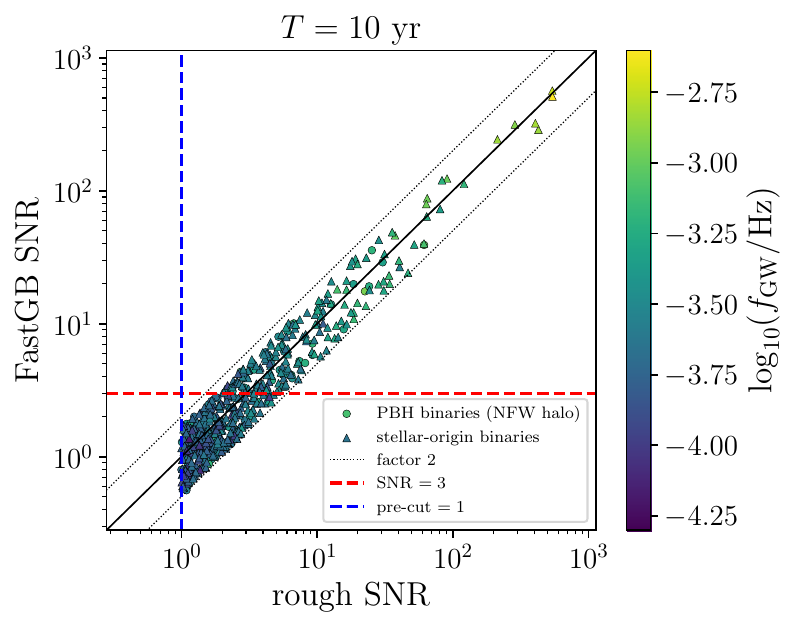}
    \caption{We compare the rough sky averaged SNR calculation with the outcome of \texttt{fastgb}, which evolves the waveform through the LISA response function over 10 years. They agree very well, with a scatter of 2 which is expected from the random orientations of the BBHs. Applying a pre-cut on the SNR of 1 is sufficient to keep sources with an SNR of $3$ after 10 years. }
    \label{fig:check_snr}
\end{figure}

We take the noise spectra in the $A,E,T$ basis from \cite{Flauger:2020qyi} with test-mass acceleration noise $A=3\times10^{-15}\,\mathrm{m\,s^{-2}\,Hz^{-1/2}}$ and
optical metrology noise $P = 15\,\mathrm{pm\,Hz^{-1/2}}$ with equal arm lengths of $L=2.5 \cdot 10^6$m. The Galactic confusion foreground of \cite{Robson:2018ifk} (Eq.\,(14)) is included as an additional stationary, isotropic noise, $S_{ij}^{\rm tot} = N_{ij} + \mathcal{R}_{ij}\, S_{\rm conf}$, with the channel noises $N_{ij}$ and responses $\mathcal{R}_{ij}$ of \cite{Flauger:2020qyi} (Eqs.\,(2.19)-(2.25)). Given two signals $a$ and $b$ in the LISA data, with Fourier transforms
$\tilde a_c(f)$ and $\tilde b_c(f)$ in the TDI channels $c\in\{A,E,T\}$,
we define the noise-weighted inner product \cite{Cornish:2007if}
\begin{equation}
  (a|b) = 4\,\mathrm{Re} \l( \sum_{c\in\{A,E,T\}} \int_{0}^{\infty} \dd f\,
  \frac{\tilde a_c(f)\,\tilde b_c^*(f)}{S_{cc}(f)} \r) \,,
\end{equation}
where $S_{cc}(f)$ is the noise power spectral density of channel $c$.
We denote by $h(\boldsymbol\theta)$ the response of LISA to a source with
parameters $\boldsymbol\theta$, i.e.\ its gravitational-wave signal in the
three TDI channels. The signal-to-noise ratio is then $\rho = (h|h)^{1/2}$,
evaluated at the true parameters of each source, including its sky position
and orientation. Sources with $\rho > \rho\e{th} = 3$ are considered detected. The comparison between the rough SNR cut based on the amplitude of the characteristic strain and the results from \texttt{fastgb} are presented in Fig.\,\ref{fig:check_snr}. The estimate of the SNR scatters around the rough estimate by a factor of $2$, which is expected from the variation in the source angle parameters. 

\begin{figure}
    \includegraphics[width=1.0\linewidth]{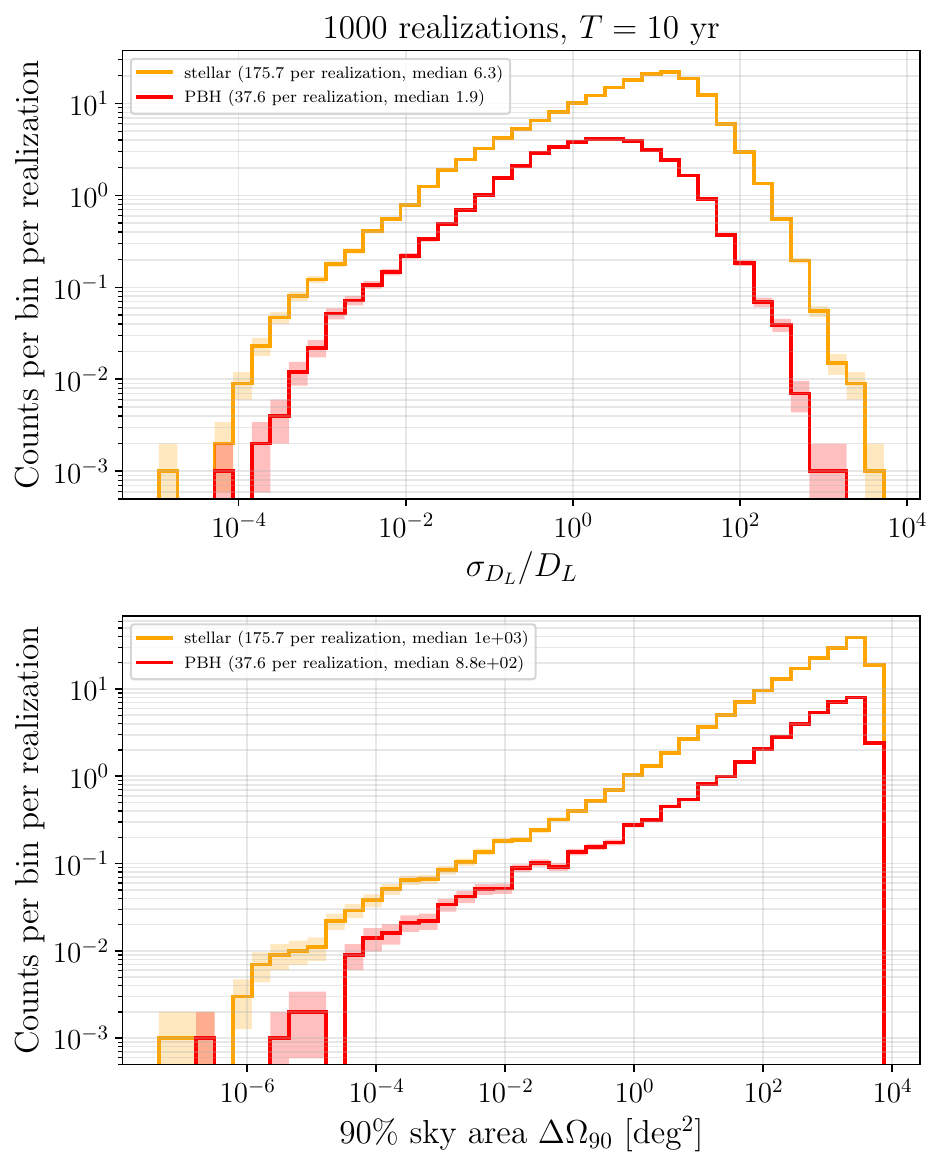}
    \caption{We present the relative uncertainties on the luminosity distance and solid angle $\Delta \Omega_{90}$ for 1000 realizations with stellar halo fraction $f\e{SH} = 0.03$.}
    \label{fig:uncertainty_histograms}
\end{figure}

For each detected source, we compute the Fisher matrix $\Gamma_{ij} = (\partial_i h|\partial_j h)$ for the eight parameters $\bs{\theta} = \{\ln\mathcal{M}, \ln D_L, f_0, \iota, \psi, \varphi_0, \lambda, \beta\}$: the logarithm of the chirp mass $\mathcal{M} = (m_1 m_2)^{3/5}/(m_1+m_2)^{1/5}$, the logarithm of the luminosity distance $D_L$, the gravitational-wave frequency $f_0$ at the start of the observation, the inclination $\iota$ of the orbital angular momentum to the line of sight, the polarization angle $\psi$, the initial gravitational-wave phase $\varphi_0$, and the ecliptic longitude $\lambda$ and latitude $\beta$ of the source. The derivative with respect to $\ln D_L$ is exact, $\partial h/\partial\ln D_L = -h$. The other derivatives are finite differences of the \texttt{fastgb} waveform. The covariance reads $\Sigma = (\Gamma)^{-1}$. Marginalizing over the other parameters amounts to keeping the $(\ln D_L, \lambda, \beta)$ block of $\Sigma$, which we transform to equatorial coordinates $(\ln D_L, \alpha, \delta)$ with the Jacobian of the ecliptic-to-equatorial rotation. The 90\% sky area is $\Delta\Omega_{90} = 2\pi\ln 10\,\cos\beta\,\sqrt{\det\Sigma_{\lambda\beta}}$ \citep{Cutler:1997ta}. We present the results for these relative uncertainties in Fig.\,\ref{fig:uncertainty_histograms}.

\begin{figure*}
    \includegraphics[width=1.0\linewidth]{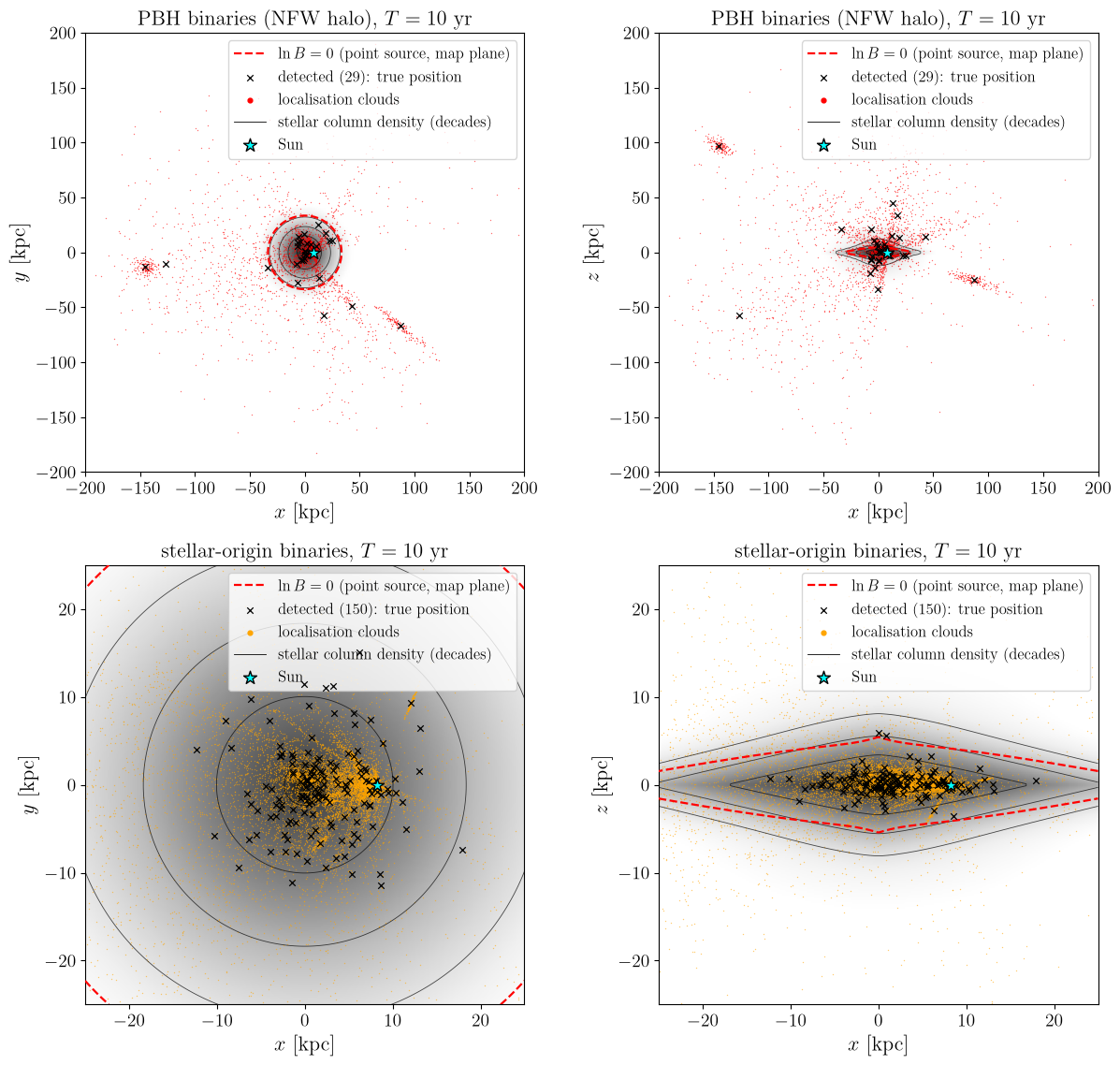}
    \caption{We present the BBHs of stellar and primordial origin originating from different regions in the Milky Way. The yellow and red clouds around the detected BBHs represent their localization volume. Note the different scales for the top and bottom plots. Some PBH binaries are localized well outside the volumes with significant stellar densities. Those are the ones for which it is easiest to use their localization volume to exclude a stellar origin. This plot is realized with a stellar halo fraction $f\e{SH} = 0.03$. The red dashed lines indicate the surface for which a point-like source would have an overlap ratio $\ln B = 0$ (see Eq.\,\eqref{eq:lnB}).}
    \label{fig:map_10yr}
\end{figure*}

\section{Stellar and NFW halo overlap} \label{sec:Stellar_Overlap}

Our goal is to determine whether a detected BBH is likely to be of stellar origin or to belong to a PBH population tracing the dark matter. Let $p\e{gw}(\bs{x})$ denote the probability density of the source position $\bs{x}$ inferred from the GW signal, which we approximate by the Gaussian distribution in $(\ln D_L, \alpha, \delta)$ given by the Fisher matrix, mapped to Galactocentric coordinates. We compute its overlaps with the stellar mass density $\rho_*$ and with the dark-matter density $\rho\e{NFW}$,
\begin{align}
Z_* &\equiv \frac{1}{M_*} \int_{\mathbb{R}^3} \dd^3\bs{x} \, p\e{gw}(\bs{x})\, \rho_*(\bs{x})\,, \label{eq:Z*} \\
Z\e{NFW} &\equiv \frac{1}{M\e{DM}}\int_{\mathbb{R}^3} \dd^3\bs{x} \, p\e{gw}(\bs{x})\, \rho\e{NFW}(\bs{x})\,, \label{eq:ZNFW}
\end{align}
and one defines the \textit{overlap ratio} as
\begin{align}
\ln B \equiv \ln \l( \frac{Z_*}{Z\e{NFW}} \r)\,. \label{eq:lnB}
\end{align}
It should be noted that while the MW stellar and DM mass appear explicitly in the overlap definitions (Eq.\,\eqref{eq:Z*} and \eqref{eq:ZNFW}), they actually cancel the masses which appear implicitly in $\rho_*$ and $\rho\e{NFW}$. This effectively normalizes the stellar and dark matter density profiles to unit integral over the Galactic volume. This removes the relative total mass normalization, $M_*/M_{\rm DM}\simeq0.052$ in the overlap ratio, which does not directly represent the relative abundance of stellar and primordial BBHs. Both densities peak at the Galactic center, so both overlaps decrease for sources further away. The stellar overlap, however, decreases faster: at large radii the disc and bulge are exponentially suppressed and only the subdominant stellar halo remains, which falls as $r^{-3.5}$, slightly more steeply than the NFW profile ($\propto r^{-3}$). Sources in the outer Galaxy therefore have $\ln B < 0$. 

Since $B$ is the ratio of two averages of the densities over the same distribution $p\e{gw}(\bs{x})$, it cannot exceed the extreme values of the density ratio,
\begin{align}
\min_{\bs{x}}\left[\frac{\rho_*(\bs{x})/M_*}
{\rho_{\rm NFW}(\bs{x})/M_{\rm DM}}\right]
\leq B\leq
\max_{\bs{x}}\left[\frac{\rho_*(\bs{x})/M_*}
{\rho_{\rm NFW}(\bs{x})/M_{\rm DM}}\right]\,,\label{eq:bounds_lnB}
\end{align}
where the extrema are taken over the dark-matter halo, bounded by $||\bs{x}|| \le 200$\,kpc. The maximum is reached in the Galactic plane at $R \simeq 0.1$\,kpc, at the edge of the bulge core, and the minimum along the poles at $r = 200$\,kpc, the edge of the dark-matter halo (see Table~\ref{tab:lnB_bounds}). Under the two normalized hypotheses, $B=1$ corresponds to equal spatial overlaps, although this does not necessarily imply equal posterior probabilities for the two formation channels, which also depend on their relative event rates and prior probabilities. A perfectly localized source in the densest stellar region reaches only $\ln B \simeq \ln(173) \simeq 5.1$. In contrast, more negative values of $\ln B$ mean that a source lives in places which have low stellar density support, such as the stellar halo or the DM halo. In the next section, we show that this parameter can be used as a way to infer how likely a GW is to originate from stellar BBH or PBH.

\begin{table}[t]
\centering
\begin{tabular}{ccc}
\toprule
$f_{\rm SH}$ & $\min_{\bs{x}} \ln B$ & $\max_{\bs{x}} \ln B$ \\
\midrule
$0.001$ & $-10.9$ & $5.1$ \\
$0.01$  & $-8.6$ & $5.1$ \\
$0.03$  & $-7.5$ & $5.1$ \\
\bottomrule
\end{tabular}
\caption{Bounds on $\ln B = \ln(Z_\star/Z_{\rm NFW})$ for different stellar halo
fractions $f_{\rm SH}$ for point-like localization. Since $B$ is a ratio of averages over the same posterior samples, it lies between the extreme values of $\rho_\star/\rho_{\rm NFW}$ within the dark matter halo ($r \le 200$\,kpc). The maximum is reached in the Galactic plane at $R \simeq 0.1$\,kpc, where the stellar halo is negligible, and is therefore independent of $f_{\rm SH}$. The minimum is reached along the poles at $r = 200$\,kpc, where the stellar density is that of the stellar halo alone, and scales as $\ln f_{\rm SH}$. Observed values lie inside these bounds, since a finite localization averages over less extreme regions.}
\label{tab:lnB_bounds}
\end{table}

\section{Results}\label{sec:Results}

\begin{figure}
    \includegraphics[width=1.0\linewidth]{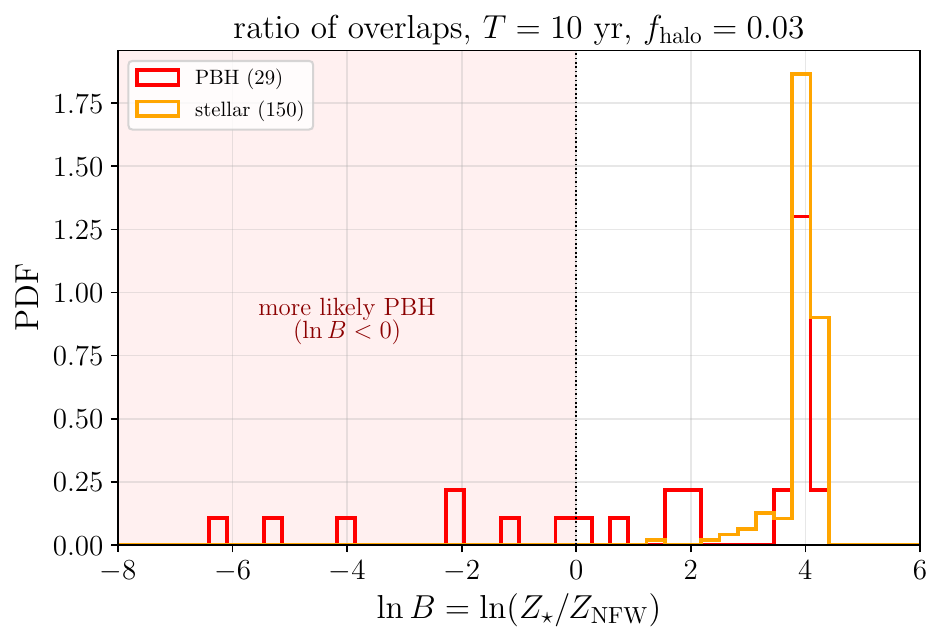}
    \caption{Distribution of stellar and PBH origin GW events in terms of their logarithmic evidence $\ln(B)$, introduced in Eq.\,\eqref{eq:lnB}. Negative values ($\ln B < 0 $) 
    indicate that the event is more likely to be from a PBH living in the NFW halo. Some PBH are indistinguishable from the stellar ones because they overlap too well with the regions of high stellar density. The discriminating GW events are those which lives well outside the stellar disc and bulge, which have low $\ln B$ as they overlap only with the stellar halo, fixed such that $f\e{SH} = 0.03$. }
    \label{fig:lnB}
\end{figure}

We now study the distribution of overlap ratio $\ln B$ in the two scenarios where all of the BBH in the MW are of stellar origin or of primordial origin. It turns out that the stellar halo provides the leading challenge in inferring the primordial origin of the BBH. Without the stellar halo, the evidence $\ln B$ for PBH can go down as low as $\sim -200$ for BBH close to the edge of the DM halo. This is due to the exponential suppression of the stellar density with galactic radius. However, the presence of the stellar halo bounds the minimum value of $\min \ln B\in [-11,-7]$ depending on the stellar halo fraction $f\e{SH} = M\e{SH}/M_*\in [0.001, 0.03]$. We present one realization of the ratio overlap for $f\e{SH} = 0.03$ in Fig.\,\ref{fig:lnB}. As predicted, the distribution of $\ln B$ is bounded to the interval $[-7.5, 5.1]$, as predicted in Tab.\,\ref{tab:lnB_bounds} from Eq.\,\eqref{eq:bounds_lnB}. Most of the 150 stellar events detected in that realization live between $2<\ln B<5$ and there are only very few events in the interval $[0,2]$ with no event below $\ln B<0$. That stems from the fact that most stellar events are localized in the disc and the bulge which is bounded due to the exponential cutoffs in densities. There are only 3\% of stellar events arising from the stellar halo which in that case are not that far out since the profile is $\rho\e{SH}(r) \propto r^{-3.5}$. In contrast, while there are many PBHs which also live in the interval $[2,5]$, there are about 7 PBH events located below $\ln B <0$, where no stellar events shows up in this realization. This suggests that the overlap ratio may provide a good diagnostic to dismiss the stellar origin of some BBH.

\begin{figure}
    \includegraphics[width=1.0\linewidth]{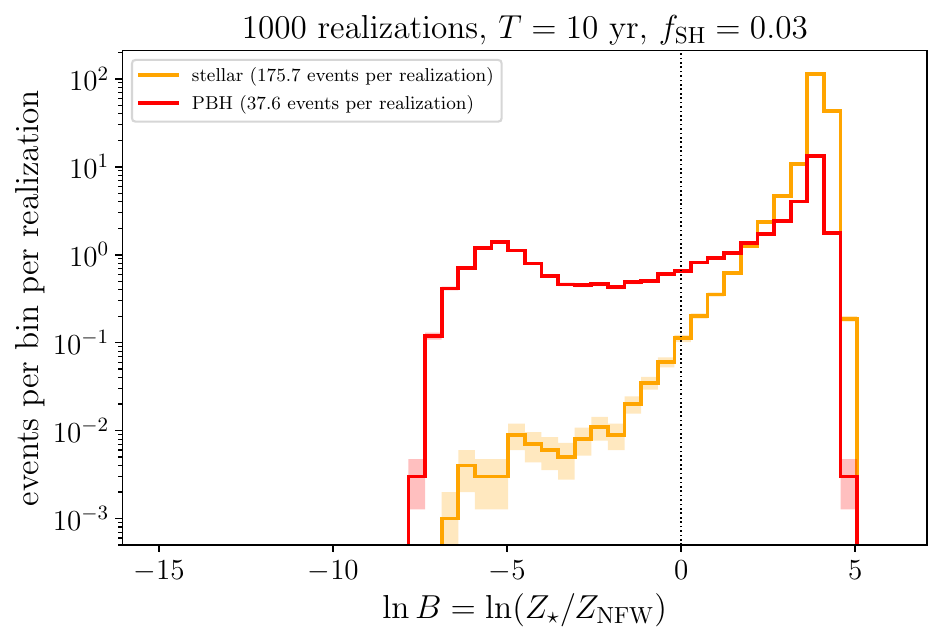}
    \caption{Distribution of stellar and PBH origin GW events in terms of their logarithmic evidence $\ln(B)$, introduced in Eq.\,\eqref{eq:lnB} for 1000 realizations with stellar halo fraction $f\e{SH} = M\e{SH}/M_*= 0.03$. In both scenarios there is a significant fraction of BBH events which have $\ln B \in [2,5]$. The events with $\ln B<0$ are most likely PBH living in the NFW halo. For stellar BBH, the number of BBH's located at smaller values of $\ln B$ decays as a power law, with a few events from the stellar halo contributing for $\ln B \in [-7.5,-3]$. For PBH's, there is also a peak of BBH at $\ln B \in [2,5]$ and the number of PBH falls with lower $\ln B$ until it reaches a second maximum close to $\ln B \simeq -5$, which corresponds to a large region of the Milky Way, where the bulge and disc no longer contribute to the stellar density. Instead, the ratio of the stellar halo and DM halo decreases slowly leading to a pile up of PBH's between $\ln B \in [-7,-4]$. On average, at least one PBH event per simulation lives in that range. GW events with negative $\ln B$ can be used to identify PBHs.}
    \label{fig:lnB_1000real_fhalo0.03}
\end{figure}

\begin{figure}
    \includegraphics[width=1.0\linewidth]{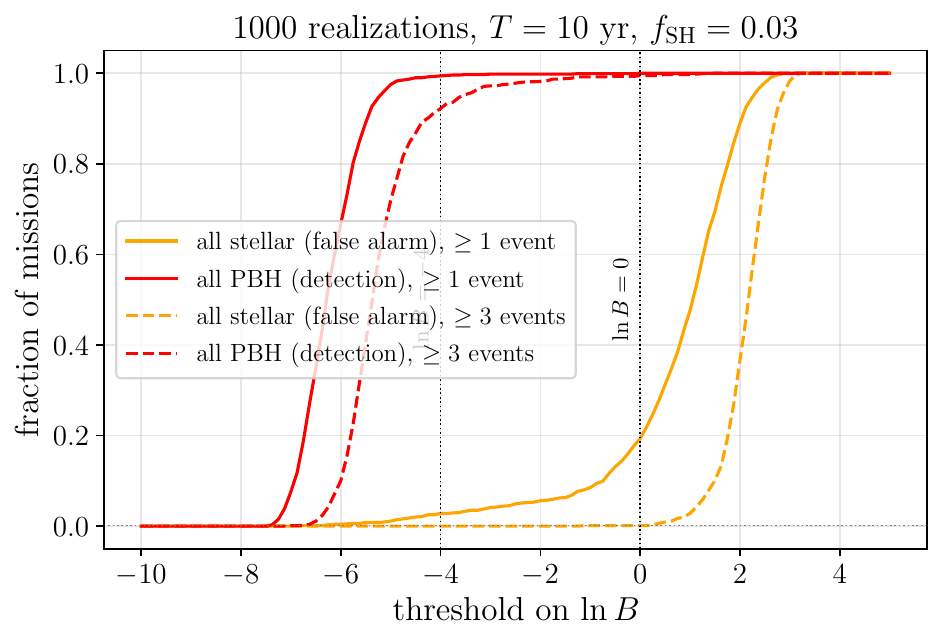}
    \caption{We present the fraction of 10 year LISA missions which detect at least 1 or 3 GW events with $\ln B$ smaller than a certain threshold ($\ln B< \ln B\e{th}$) in the two scenarios, where all BBH are stellar or primordial in origin for $1000$ realizations. For example, in the PBH scenario, it is very likely that one detects at least on GW with $\ln B \leq -4$. This happens in $99\%$ of simulations, while this is quite unlikely for stellar stellar BBH for which it happens in only $3\%$ of simulations. This is perhaps even more exciting for the detection of more than 3 GWs with $\ln B\leq 0$, which happens in $99\%$ of PBH simulations and in less than $0.1\%$ of stellar BBH simulations. }
    \label{fig:mission_test_10yr_fhalo0.03}
\end{figure}

Next, we run $1000$ simulations to quantify more precisely how common it is for PBH versus stellar BBH to populate the negative region of the overlap ratio. We present the average distribution of $\ln B$ in Fig.\,\ref{fig:lnB_1000real_fhalo0.03}. There, one may notice that the distribution of $\ln B$ falls essentially as a power law towards negative values of $\ln B$ until $\ln B = -5$. There, the contribution from the stellar halo ($\mathcal{O}(10^{-2})$ of events) kicks in. In contrast, the PBH distribution also falls with negative values of $\ln B$ until it reaches a minimum around $\ln B = -3$ and then a maximum around $\ln B \simeq -5$. This peak comes from the fact that the density ratios between stars and DM falls steeply at low radius until the edge of the stellar disc. Beyond that, the ratio of densities falls as $r^{-0.5}$ since the stellar halo starts to dominate the stellar density. Hence, there is a large region of the Milky Way between $r\in[10,200]$kpc, where $\ln B$ evolves very slowly towards the lower bound and leads to a pile up of events between $-7.5<\ln B<-3$. Note that there is on average, at least $1$ PBH with $\ln B < -4$ in each realization. This histogram can be used to quantify how likely it is to detect a certain number of BBH events below a threshold, i.e.\,$\ln B< \ln B\e{th}$. We present the fraction of realizations in which the observer detects more than 1 (or 3) BBH events with an overlap smaller than the threshold, as a function of the threshold in Fig.\,\ref{fig:mission_test_10yr_fhalo0.03} for the two scenarios. This allows to identify that detecting at least one BBH with $\ln B\leq -4$ happens in $99\%$ of simulations if the BBH are of primordial origin. In contrast, it happens in only $3\%$ of simulations if the BBH are of stellar origin. This suggests that values of $\ln B < -4$ may be used as evidence of primordial origin for BBH, although this is not excluded for a stellar halo origin BBH. For more confidence, detecting 3 BBH with $\ln B<0$ happens in $99\%$ of simulations if the BBH are primordial, while it happens in less than $0.1\%$ \, of simulations if the BBH are of stellar origin. Therefore, only as few as 3 detections of MW BBH localized in the halo would be sufficient to conclude confidently that the BBH are of primordial origin. As a reminder, for a point-like source, $\ln B = 0$ denotes a surface surrounding the stellar disc which is represented in Fig.\,\ref{fig:map_10yr} in red-dashed. A source well outside of the enclosed volume would have $\ln B<0$. 

\begin{figure}
    \includegraphics[width=1.0\linewidth]{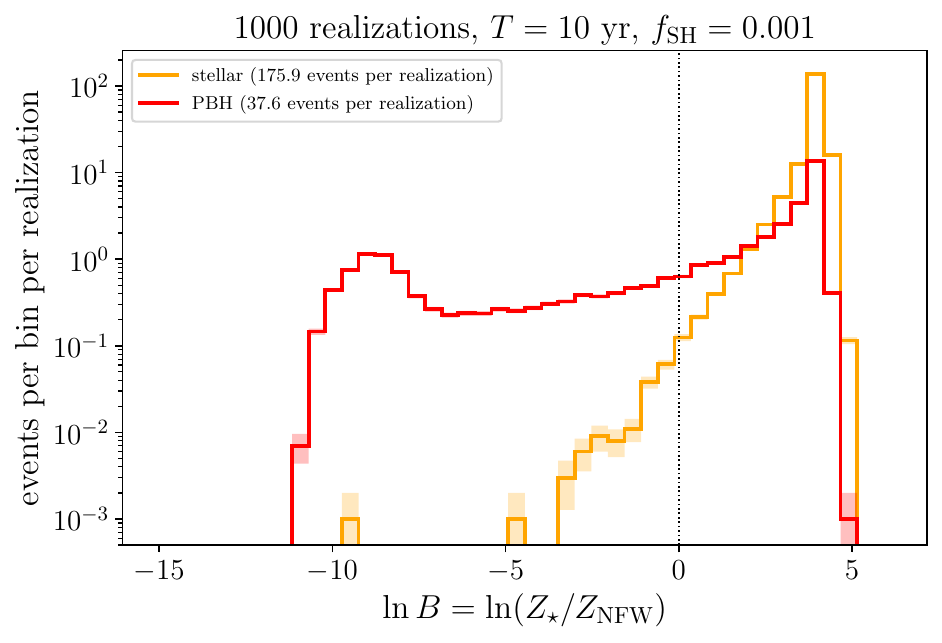}
    \caption{Distribution of stellar and PBH origin GW events in terms of their logarithmic evidence $\ln(B)$, introduced in Eq.\,\eqref{eq:lnB} for 1000 realizations with a relatively low stellar halo fraction $f\e{SH} = M\e{SH}/M_*= 0.001$. There is a broader range ($\ln B \in [-11,-4]$) of negative values for PBHs where there are very little stellar BBH compared to Fig.\, \ref{fig:lnB_1000real_fhalo0.03}.}
    \label{fig:lnB_1000real_fhalo0.001}
\end{figure}

\begin{figure}
    \includegraphics[width=1.0\linewidth]{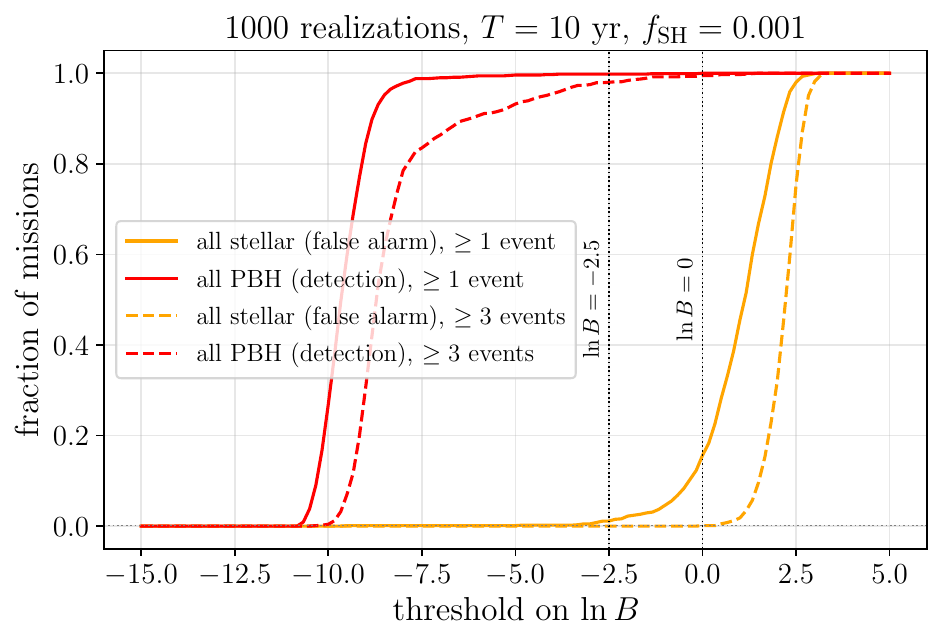}
    \caption{For a stellar fraction $f\e{SH} = M\e{SH}/M_* =0.001$, we present the fraction of 10 year LISA missions which detect at least 1 or 3 GW events with $\ln B$ smaller than a certain threshold in the two scenarios, where all BBH are stellar or primordial in origin for $1000$ realizations. Three GW event with $\ln B\leq 0$, which happens in $99\%$ of PBH simulations and in less than $0.1\%$ of stellar BBH simulations. There is at least on PBH with $\ln B\leq -2.5$ in more than $99.9\%$  of simulations, while this happens in $1\%$ of stellar BBH simulations.}
    \label{fig:mission_test_10yr_fhalo0.001}
\end{figure}

We also show the distribution of ratio overlaps for a more optimistic stellar halo fractions in Fig.\,\ref{fig:lnB_1000real_fhalo0.001} for $f\e{SH}= 0.001$. There one can see that the PBH ratio overlaps extend even further into the negative region all the way to $\ln B =-10.9$, while stellar BBH are rarer in the interval $-11<\ln B<-4$ than for $f\e{SH} = 0.03$. We also present the fraction of missions for which one detects 1 (or 3) GW events with $\ln B$ smaller than a threshold in Fig.\,\ref{fig:mission_test_10yr_fhalo0.001}. There, one can see that the threshold region where there is a significant difference between stellar BBH and PBH is wider $\ln B\e{th} \in[-7.5,-2.5]$. The interesting upper boundary $\ln B <0$ could even be pushed to $0.5$ for 3 GW detections. However, to be conservative, one may keep $\ln B\e{th} = 0$ as a robust threshold for the inference of the existence of PBH which contributes to DM. There, detecting an event with $\ln B < -2.5$ happens in more than $99.9\%$ of simulations, while it happens in $1\%$ of stellar BBH simulations. 

These findings demonstrate that finding evidence for a primordial origin of BBH in the Milky Way based on their 3D localization is feasible, while it depends on the effective galactic merger rate. It provides complementary information to the other means of discriminating PBH from stellar ones based on other parameters such as mass and spin.

\section{Conclusion}\label{sec:Conclusion}

In this work, we studied how the three-dimensional localization of gravitational-wave (GW) signals from Galactic binary black holes (BBHs) with LISA can help distinguish between stellar and primordial origins. Our method exploits the different spatial distributions expected for these populations: stellar-origin BBHs are associated with the stellar components of the Milky Way, whereas primordial black holes (PBHs) contributing to dark matter may trace its more extended halo. This provides a complementary diagnostic to those based on the masses, spins and orbital properties of the binaries.

We used an optimistic merger rate of BBH $R_{\rm MW}=3\times10^{-5}\,{\rm yr}^{-1}$, motivated by current LVK measurements, to predict the number of BBH inside the Milky Way, below a certain semi-major axis separation threshold. We found that there could be as many as $1.5\times10^5$ BBHs with separations below $10R_\odot$ and component masses in the interval $[4,100]M_\odot$, under our adopted mass distribution and merger-rate normalization.

We performed simulations and distributed these BBH inside the Milky Way, following a simple matter density model. We performed two sets of simulations. In the first scenario, these BBH were assumed to be
of stellar origin, distributed following the stellar mass density, including the Galactic bulge, thick disc and stellar halo. In a second scenario, these BBH were assumed to be of primordial origin and distributed according to an NFW profile, modeling the DM component of the Milky Way, where the primordial BBH are expected to contribute. We simulated the internal properties of these BBH such as masses and inclination and used \texttt{fastgb} to determine the SNR of these BBH
in the LISA band. We performed a Fisher forecast to determine the uncertainties on each of these parameters, with a special focus on the three-dimensional localization of the sources. We approximated the resulting localization posteriors as Gaussian distributions, an assumption that may become inaccurate for sources near the
detection threshold or with multi-modal posteriors.

We introduced the overlap ratio $\ln B$ to quantify how well each of these BBH overlaps with the stellar density and DM profiles. The ratio compares the overlap of each GW localization posterior with the unit-normalized stellar and dark matter density profiles, and therefore depends on their spatial distributions rather than their relative total masses. We found that most BBH of stellar and primordial origin have a positive overlap ratio which does not allow to discriminate their origin. Indeed, most detected GW events happen in the vicinity of the Galactic center and disc, which have significant overlap with both
the DM and stellar profiles. However, there is a long tail of PBH events which have $\ln B<0$, while there are only very few stellar-origin BBHs with such values, primarily associated with the stellar halo.

We ran 1000 realizations to quantify the probability of detecting one or several events below a given overlap threshold, under
the stellar and primordial hypotheses. We found that detecting at least one event with $\ln B<-4$ occurs in approximately 99\% of the PBH simulations, compared with about 3\% of the stellar-origin simulations for $f_{\rm SH}=0.03$. More significantly, at least three detections with $\ln B<0$ occur in approximately 99\% of the PBH simulations, while no such cases were found among the 1000 stellar-only realizations. This corresponds to an empirical false-alarm frequency below $10^{-3}$ at the resolution of our simulations, although the statistical uncertainty implies an approximate 95\% upper limit of $3\times10^{-3}$ on the underlying probability. Lowering the assumed stellar-halo fraction from $f_{\rm SH}=0.03$ to $0.001$ further improves the separation between the two populations. These results demonstrate that the localization of only a few Galactic BBHs could provide strong evidence against a conventional stellar-origin population, within the assumptions of our model.

We have assumed a MW merger rate of $3\times10^{-5}\,{\rm yr}^{-1}$, towards the optimistic end of our estimates based on current LVK measurements. Lower Galactic rates would reduce the number of LISA detections and consequently the probability of obtaining several discriminating events. Nonetheless, predicting the MW merger rate with current data is difficult. This is because the translation from the cosmic average rate to the Milky Way rate is nontrivial and may depend on the formation channel of these BBH. Additionally, there may be a complicated dependence of the merger rate on metallicity, star formation history and galactic merger history of the Milky Way \cite{Lamberts:2018}. In particular, the merger rates of stellar-origin and primordial BBHs need not be equal, and the detection probabilities presented here should be interpreted as conditional on the adopted population normalizations.

It should also be mentioned that we have assumed a simple MW mass density model to easily sample the BBH positions from
the stellar or DM density. However, the methodology is not bound
to this mass model and one could imagine using a more sophisticated analysis to predict the stellar BBH population in the Milky Way as in \cite{Lamberts:2018}. The most important astrophysical
uncertainty is the distribution of stellar-origin BBHs in the Galactic halo, whose abundance need not simply follow the stellar mass fraction. Metallicity, binary evolution, dynamical formation and natal kicks could all affect their spatial distribution. Likewise, primordial BBH populations need not trace the NFW profile exactly, depending on their
formation and dynamical history. The output of more realistic population synthesis models and Galactic simulations may be used to refine the estimates presented in this work. We have also not dealt with extracting the BBH signals from a noisy LISA strain, beyond the expected galactic White Dwarfs, which represents a research topic of its
own \cite{Cornish:2005qw}. A full Bayesian analysis incorporating realistic localization posteriors, selection effects and the relative rates of the two populations would provide a natural extension of this work.

Finally, we conclude that LISA's ability to localize Galactic BBHs
in three dimensions opens a promising new avenue for investigating their origin. Even a small number of binaries localized in regions where the stellar population is strongly suppressed could provide compelling evidence for a population tracing the Galactic dark matter halo. This spatial diagnostic is independent of, and complementary to, traditional gravitational-wave tests of primordial black holes, offering a new opportunity to probe their possible contribution to dark matter.

\vspace{5mm}

\noindent{\bf Acknowledgements} 
The authors would like to thank Lina Necib and Julien Lavalle for useful conversations at an early stage of this work. This work made use of the Python packages
NumPy~\cite{Harris:2020xlr}, SciPy~\cite{Virtanen:2019joe}, pandas~\cite{McKinney:2010}, Matplotlib~\cite{Hunter:2007}, \texttt{fastgb}~\cite{fastgb, Cornish:2007if}, and \texttt{lisaorbits}~\cite{Martens:2021phh}. EG acknowledge grants from MICINN/MICIU/AEI (/10.13039/501100011033) under grant PID2024-156844NB-C21.

\begin{appendix}

\end{appendix}

\bibliographystyle{apsrev4-2}
\bibliography{references} 

\end{document}